\documentclass[twocolumn,prl,showpacs,preprintnumbers,groupedaddress]{revtex4}
\usepackage{amssymb}
\usepackage{mathptm}
\usepackage{dcolumn}
\usepackage{bm}
\usepackage{times}
\usepackage{graphicx}

\newcommand {\be}{\begin{equation}}
\newcommand {\ee}{\end{equation}}
\newcommand {\bea}{\begin{eqnarray}}
\newcommand {\eea}{\end{eqnarray}}

\begin{document}

\title{Possible phase-sensitive tests of pairing symmetry in pnictide
superconductors}
\author{D.~Parker$^{1}$}
\author{I.I.~Mazin$^{1}$}
\affiliation{$^1$Naval Research Laboratory, 4555 Overlook Ave. SW, Washington, DC 20375}
\date{\today}

\begin{abstract}
The discovery of the new class of pnictide superconductors has engendered a
controversy about their pairing symmetry, with proposals ranging from an
extended s-wave or ``s$_{\pm}$'' symmetry to nodal or nodeless d-wave
symmetry to still more exotic order parameters such as p-wave. In this
paper, building on the earlier, similar work performed for the cuprates, we
propose several phase-sensitive Josephson interferometry experiments, each
of which may allow resolution of the issue.
\end{abstract}

\maketitle

Identification of order parameter symmetry is one of the first tasks the
condensed matter physics community faces upon discovery of a new
superconductor. Historically, as pointed out by Van Harlingen \cite{harlingen},
methods of determining order parameter symmetry have fallen into two
classes: techniques which are sensitive to the \textit{magnitude} of the
order parameter, and techniques which are sensitive to the \textit{phase}.
Most of the magnitude sensitive techniques are ultimately concerned with the
presence of Fermi surface nodes. Examples include thermodynamic tests such
as density-of-states, specific heat, and London penetration depth. The first
experimental technique to yield detailed information about the momentum
dependence of the order parameter was angle-resolved photoemission
spectroscopy (for a review, see Ref. \cite{shen}), or ARPES, which
demonstrated the substantial momentum anisotropy in the high-temperature
cuprate gap function.

None of these tests, however, is a \textquotedblleft smoking
gun\textquotedblright\ ultimately capable of unequivocally determining the
order parameter structure. For this one also requires a phase-sensitive
test, such as the Josephson interferometry \cite{wollman} or tricrystal
junctions \cite{tsuei}. Such tests, as originally proposed by Geshkenbein et
al \cite{geshkenbein}, Rice and Sigrist \cite{rice}, and Leggett \cite%
{wollman} provided highly convincing evidence for d-wave superconductivity
in the cuprates, effectively ending a controversy of several years, and have
been also used to address p-wave superconductivity in Sr$_{2}$RuO$_{4}$\cite%
{nelson}.

We now consider such a test of order parameter symmetry in the pnictide
superconductors, which have been extensively investigated since the original
discovery by Kamihara early in 2008 \cite{kamihara}. There are now dozens of
superconductors in this family, with superconducting transition temperatures 
$T_{c}$ as high as 57 K. Bandstructure calculations and ARPES data indicate
that these materials contain disjoint Fermi surfaces, as illustrated in
Figure 1, with a hole pocket 
\begin{figure}[h]
\includegraphics[width=7.5cm,angle=0]{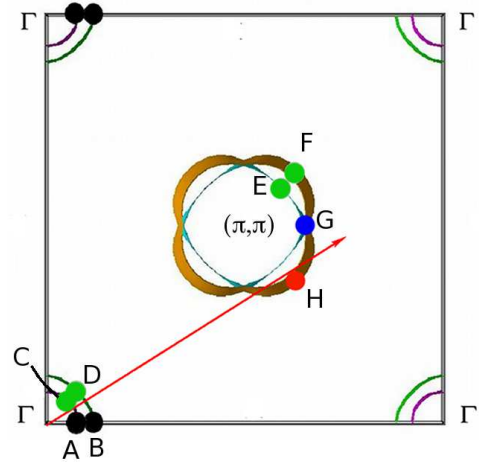} \vspace{-0.7cm}
\caption{(Color online) A view of the calculated Fermi surface geometry in a
superconducting pnictide LaFeAsO$_{0.9}$F$_{0.1}$, with hole ($\Gamma $) and
electron pockets ($\protect\pi,\protect\pi$) indicated. For a thick barrier
the black circles represent the Fermi surface states which dominate the
(100) current, while the green circles represent the states which dominate
the (110) current. A possible intermediate angle, where the electron surface
may dominate the current, is shown by the arrow. Greek characters represent
standard BZ points, while Roman characters refer to the adjacent circles
whose wavefunction character is given in Table 1.}
\end{figure}
centered around (0,0) and electron pockets at $(\pi ,\pi )$ and related
points.

Despite this effort, the gap symmetry of the pnictides remains unknown. A
potential pnictide gap function presently receiving much consideration is
the \textquotedblleft s$_{\pm }$\textquotedblright\ state \cite{mazin}, in
which the order parameter changes sign from the hole to electron Fermi
surfaces, but is roughly constant on each Fermi surface, with no nodes.

To date, there have been three phase-sensitive experiments performed on the
pnictides. The first is the observation in inelastic neutron scattering
(INS) measurements on Ba$_{0.6}$K$_{0.4}$Fe$_{2}$As$_{2}$\cite{christianson}
of a resonance peak centered at $\mathbf{Q}=(\pi ,\pi $) that appears below $%
T_{c}$. This effect has been well-studied in connection to the cuprates \cite%
{demler_rmp}, and in pnictides it had been predicted theoretically for the $%
s_{\pm }$ states because of the change in order parameter sign \cite%
{mazin,scalapino,eremin_epl} over the vector \textbf{Q}. More recently, an
ab-corner junction experiment was performed \cite{zhou} on BaFe$_{1.8}$Co$%
_{0.2}$As$_{2}$, which found no evidence for a phase shift between the a and
b directions, suggesting that the d-wave symmetry observed in the cuprates
is not present in this material. Similarly, Zhang et al \cite{zhang}
fabricated c-axis Josephson junctions between a conventional superconductor
and Ba$_{1-x}$K$_{x}$Fe$_{2}$As$_{2}$ and observed Josephson coupling,
suggestive of an s-wave state, but not providing clear evidence for the s$%
_{\pm}$ state itself.

In this paper we propose direct phase-sensitive tests, based upon Josephson
interferometry, that could provide strong evidence for an s$_{\pm }$ state,
if existent. The proposal is based on an adaptation of the famous
\textquotedblleft corner junction" experiments performed for the cuprates.

We briefly review the theory of corner junctions and their application to
the cuprates and Sr$_{2}$RuO$_{4}$. In a corner junction, the Josephson
current is allowed to flow from two separate faces of a single crystal of
unconventional superconductor. A junction usually preferentially samples
current oriented along the normal to the interface. By measuring the
critical current flow as a function of magnetic field, one can determine the
phase difference between the two directions sampled. Such experiments were
enormously successful in determining the pairing symmetry in the
high-temperature cuprates \cite{wollman}, and have been also applied to Sr$%
_{2}$RuO$_{4}$.\cite{nelson}

One key to these experiments has been the existence of symmetry constraints
dictating a particular phase difference for specific crystallographic
directions. In d-wave superconductors, the phase must change by $\pi $ upon
a 90$^{o}$ rotation, while in p-wave materials upon a 180$^{o}$ rotation.
For an s$_{\pm }$ state, as presumed in the pnictides, the situation is more
complicated. No combination of tunneling directions would provide the
desired phase difference by symmetry. The a and b directions are strictly
equivalent. One has to look for two \textit{inequivalent} directions such
that one will be quantitatively dominated by hole and the other by electron
bands. In the simplest approximation of a specular (infinitely thin) barrier
and constant matrix elements this amounts to comparing the number of
conductivity channels for each direction, given by the DOS-weighted average
of the corresponding Fermi velocity, $e.g.,$ $n_{z}=\left\langle
N(E_{F})v_{Fz}\right\rangle $\cite{MazinPRL}. Unfortunately, one realizes
right away that in the e-doped compounds transport in all directions
(including $c)$ is dominated by the e-pocket \cite{mazin} (cf. Figure 1),
and in the hole doped by holes (Figure 4, dark red). Thus, phase-sensitive
experiments do not at first appear to be feasible for detecting an s$_{\pm }$
state in the pnictides. \newline

This is however no longer true for a barrier of an appreciable thickness.
While for a specular barrier, all wavevectors from all Fermi surfaces
contribute to the current, regardless of tunneling direction, for a thick
barrier electons tunelling normally to the interface have an exponentially
big advantage over those with a finite momentum parallel to the interface, $%
k_{\parallel }\neq 0.$ For instance, the tunneling probability $T_{\mathbf{k}}
$ for a simple vacuum barrier can be expressed as\cite{mazin_epl} 
\bea
T_{\mathbf{k}}=
\hspace{0.9 \columnwidth}&\nonumber\\
\frac{4m_{0}^{2}\hbar ^{2}K^{2}v_{L}v_{R}}
{\hbar
^{2}m_{0}^{2}K^{2}(v_{L}+v_{R})^{2}+ (\hbar
^{2}K^{2}+m_{0}^{2}v_{L}^{2})(\hbar ^{2}K^{2}+m_{0}^{2}v_{R}^{2})\sinh
^{2}(dK) }&&\nonumber
\eea
Here $m_{0}$ is the electron mass, $v_{L,R}$ are the Fermi velocity
projections on the tunneling directions, $d$ is the width of the barrier,
and the quasimomentum of the evanescent wavefunction in the barrier, $iK,$
is, from energy conservation,%
\[
K=\sqrt{k_{\parallel }^{2}+2(U-E)m_{0}},
\]%
where $U$ is the barrier height. The above formula is an immediate
asymmetric generalization of the textbook result \cite{landau}. Similarly to
the known result, this formula does not account for the variation of the
tunneling matrix elements due to the symmetry of actual electronic states,
which, as discussed later, may be important.

So let us for the moment concentrate on thick barriers. Note that a thick
barrier need not have very low transparency: the transparency is defined by
both height (which may be low) and thickness, while the filtering properties
are defined by the thickness only. 

Obviously, for tunneling along the (100) direction the hole transport will
fully dominate, as the electron Fermi surfaces will have a huge $%
k_{\parallel }$ of approximately $\pi /a$, with $a$ the lattice constant,
and will be exponentially suppressed. So, for a thick low barrier the (100)
Josephson current will be dominated by the hole states, while for the (110)
direction both holes and electrons will contribute (all Fermi surfaces will
have points with $k_{\parallel }=0,$ cf. Fig. 1). 

However, as is well known in the theory of spin-polarized tunneling,
occasionally tunneling from the zone center ($k_{\parallel }=0)$ is
forbidden by symmetry and the current procceds through \textquotedblleft hot
spots\textquotedblright\ with some finite $k_{\parallel }$ and is
correspondingly suppressed \cite{mazin_epl}. This depends critically upon
the character of the wavefunctions on the Fermi surface, for the
corresponding $k$ direction. So let us see how the symmetry of the wave
function will affect the tunneling matrix elements in pnictides for
different directions. Some calculated\cite{calc} orbital characters are
listed in the Table. First of all, we observe that  for the (100) direction
two hole bands contribute (points C and D).  They have wavefunctions of
primarily $xy/yz$ character, with considerable admixture of $z^{2}$ and $xy$
states (this is allowed because despite a tetragonal symmetry the $z=0$
plane is not a mirror plane in the real space). The $xz/yz$ orbitals are odd
with respect to $z\rightarrow -z$ reflection, so one can expect tunneling
from these orbitals to be suppressed for a thick vacuum (and most other)
barriers. Thus the Josephson current for the holes will be mostly controlled
by the relative admixture of the $z^{2}$ character, and, except for the 100
direction (because the $xy$ orbital is odd with respect to $x\rightarrow -x),
$ of the $xy$ character. On the other hand, the electron pockets are mostly
made up by the $xz/yz$ and $x^{2}-y^{2}$ character. Again due to their
parity neither of this orbital can tunnel exactly at direction (110)
(because $x^{2}-y^{2}$ is odd with respect to the $x\rightarrow y$
reflection). Thus in both (100) and (110) directions the current will be
dominated by holes.

But all is not lost. For an in-plane direction deviating from (110) by an
angle $\alpha ,$ the tunneling from the hole pocket $xz/yz$ orbitals will still be
suppressed, while that from the electron pocket $x^{2}-y^{2}$ orbital will only be weakened by a factor of $%
\sin ^{2}2\alpha .$ The maximum $\alpha $ at which the electron Fermi
surface still crosses the $k_{\parallel }=0$ line in the Brollouin zone
corresponds to the line $\Gamma $F$^{\prime }$ in Fig. 1; for the 10\%
e-doping, as shown  in the Figure, the $\alpha _{\max }\approx 15^{o},$ $%
\sin ^{2}(2\alpha _{\max })\approx 1/4.$ Of course, exactly at $\alpha
_{\max }$ the Fermi velocity has zero normal component so that the optimal $%
\alpha $ is close to $\alpha _{\max }$ but smaller. A back-of-the-envelope
estimate tells us that the optimal $\alpha $ is about ($3/4)\alpha _{\max }$
and that the Fermi velocity factor for that $\alpha $ suppresses the current
by a factor of two, roughly. The total factor is ($v_{\perp }/v)\sin
^{2}(2\alpha _{opt})\approx 0.1.$ From the Table, we can estimate the
corresponding factor for tunneling from the hole bands to be 0.2-0.3 (adding
up the $xy$ and the $z^{2}$ and accounting for the angular dependence).  However, the Fermi velocity
(from the first principles calculations) is greater near the electron-surface H point than the hole-surface D point by a factor of approximately 2.5, so the overall factors are roughly equal.

According to this rough estimate, the holes and electrons contribute equally to the near-(110) current,
making problematic the observation of a Josephson $\pi $-contact pair. However, one should
remember that all estimates above are very crude, order of magnitude
estimates that neglect a number of factors, such as the possibility of a larger
superconducting gap for the electron Fermi surface or, most importantly,
detailed (unknown) characteristics of the contact. We conclude that there is still
some chance of observing a $\pi$ phase shift in this experiment, and this geometry is still worth pursuing. Importantly, we can
say that the optimal angle between two interfaces should be $\sim 30-35^{0}$.  In addition, a more strongly electron-doped pnictide would tend to enlarge both the electron/hole Fermi velocity ratio and angle $\alpha_{\max}$, increasing the chance of the electron Fermi surface dominating the near-(110) current.

\begin{table}[tbp]
\caption{First-principles orbital band character from Fig. 1}
\begin{center}
\begin{tabular}{|c|c|c|c|c|c|c|c|}
\hline
\multicolumn{8}{|c|}{\bfseries Wavefunction Character} \\ \hline
\itshape Point & A & B & C & D & E & F & G \\ \hline
$xz/yz$ & 0.879 & 0.717 & 1.0 & 0.724 & 0.921 & 0.003 & 0.869 \\ \hline
$x^{2}-y^{2}$ & 0.0 & 0.0 & 0.0 & 0.0 & 0.079 & 0.997 & 0.130 \\ \hline
$z^{2}$ & 0.121 & 0.0 & 0.0 & 0.069 & 0.0 & 0.0 & 0.001 \\ \hline
$xy$ & 0.0 & 0.282 & 0.0 & 0.207 & 0.0 & 0.0 & 0.0 \\ \hline
\end{tabular}%
\end{center}
\end{table}

Fortunately, one can think of some more promising designs. Indeed, let
us consider a corner-junction experiment where the (100) junction is a
thick-barrier contact (which as we just discussed, is dominated by the
h-pockets), and the second contact is a (010) or a (001) \textit{specular}%
-barrier junction. As discussed in the beginning, either of these last
contacts in an electron-doped material will be dominated by the e-pockets,
thus providing the desired $\pi $ shift. A possible geometry is illustrated
in Figure 2, if the $s_{\pm }$ state is present.

The basic point here is that, unlike in the cuprates and Sr$_{2}$RuO$_{4}$,
directional selection is not sufficient to select the appropriate region of
Fermi surface to sample to uncover a $\pi $ phase shift. One must use
additional selection means, in this case given by the use of different
barrier characteristics in different directions.

Regarding the width and height of the non-specular potential barrier, the
key consideration is that the electron FS be suppressed greatly without a
comparable suppression of the hole FS. For a moderate barrier height $%
U-E=0.25$ eV (which would require a barrier made out of a small-gap
semiconductor, $E_{g}\sim 0.5$ eV) a barrier of width 20 \AA\ would only
suppress the hole-like Fermi surface by roughly a factor of 9 ($\sinh
^{2}1.8 $), while suppressing the electron Fermi surface by a factor of $%
\sinh ^{2}20\sim 10^{16}.$ The calculated hole-like surface suppression
factor neglects the effect of a small but finite range in k$_{\parallel}$
for this Fermi surface, whose inclusion could result in a somewhat larger
suppression. We also implicitly assume a substantially electron-doped
pnictide, so that specular transport is governed uniquely by the electron
Fermi surface. 
\begin{figure}[h]
{\normalsize \includegraphics[width=3cm,angle=90]{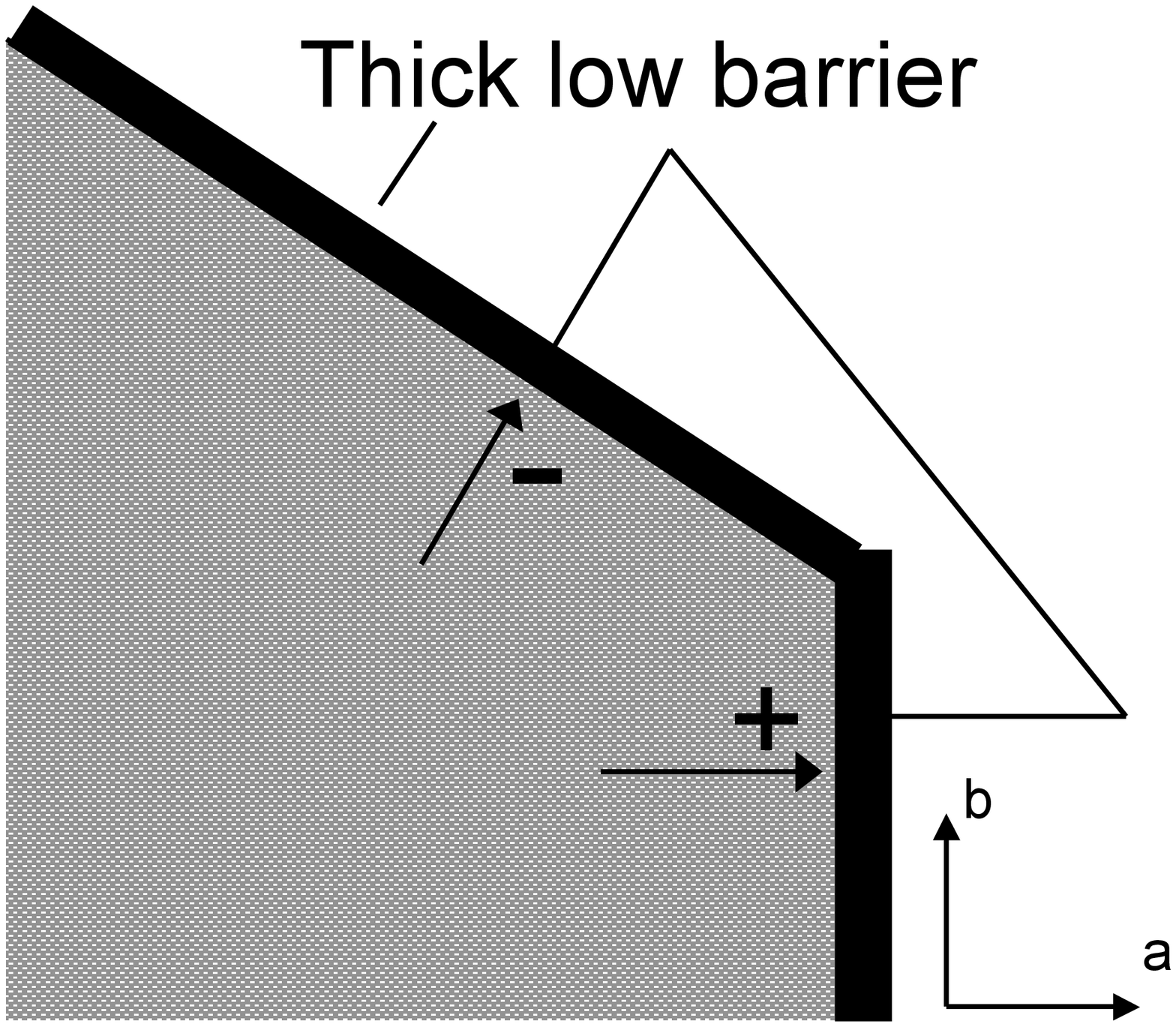}  %
\includegraphics[width=3.75cm,angle=0]{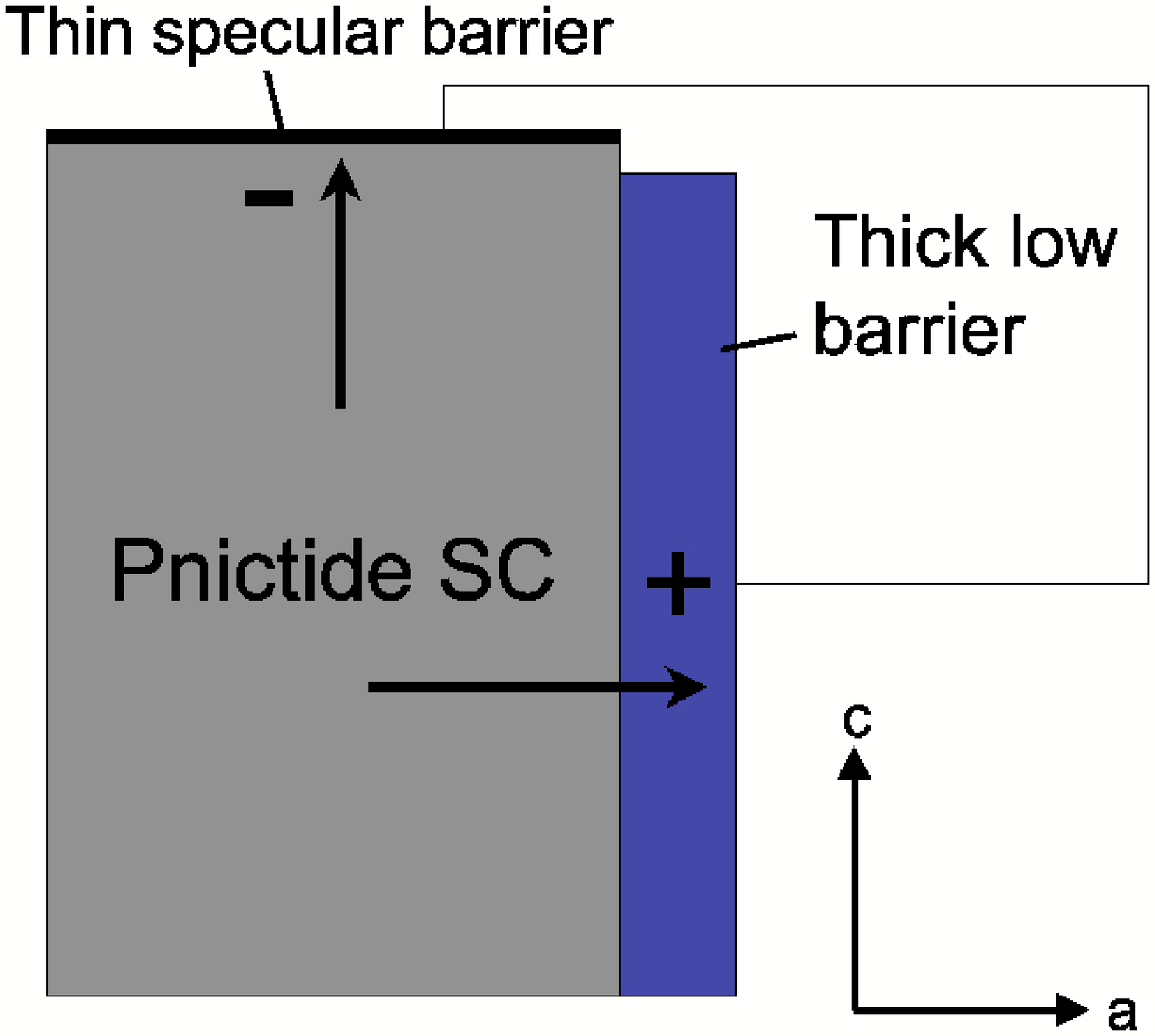}  } 
\caption{A schematic view of tunneling geometry for two possible
experiments: left, a (100) -near-(110) orientation, right, an ac orientation
with specular and thick barriers as indicated.}
\end{figure}
\vspace{-0.4cm}  
\begin{figure}[h!]
{\normalsize \includegraphics[width=3.75cm,angle=270]{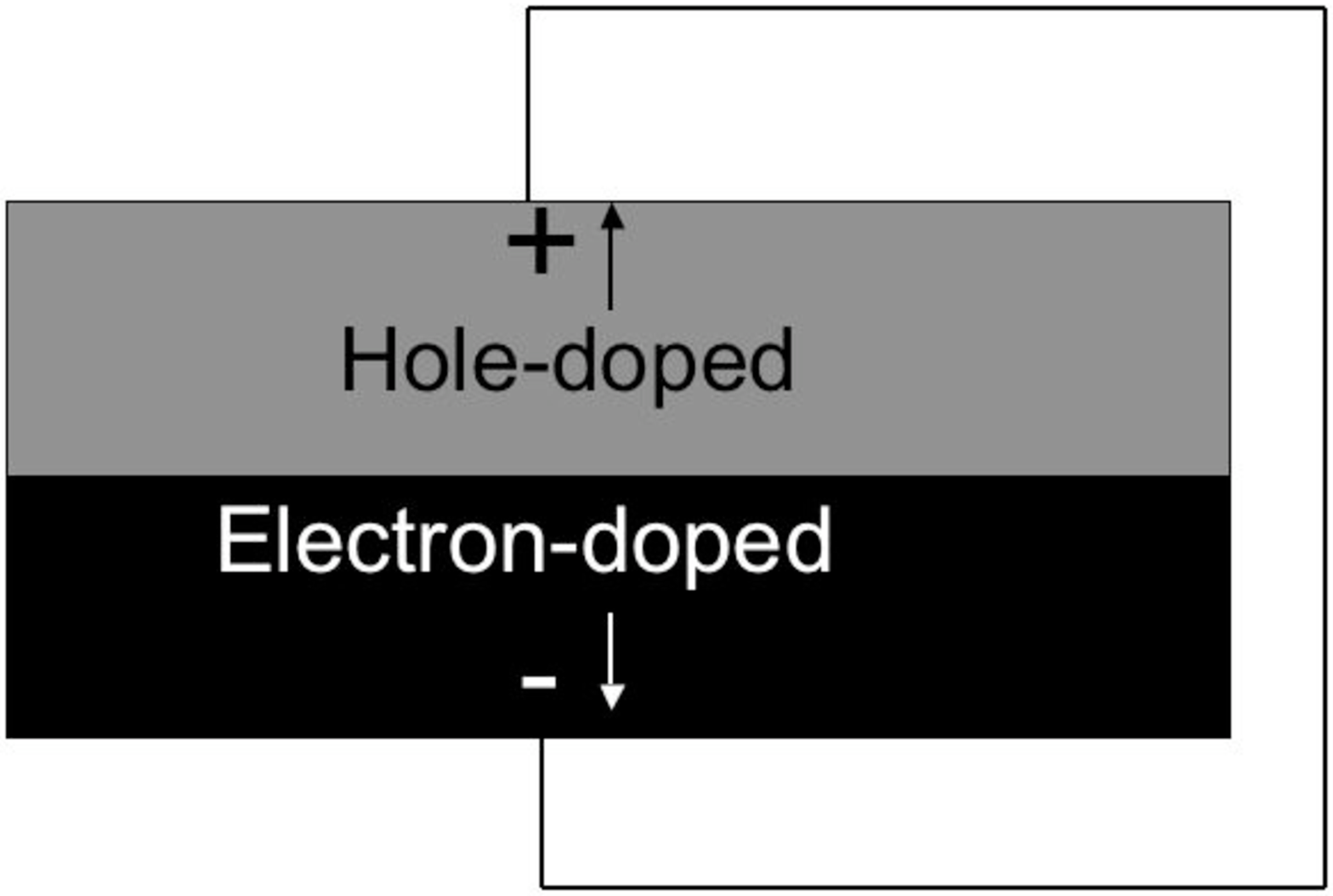} 
\vspace{-0.7cm} \includegraphics[width=4cm,angle=90]{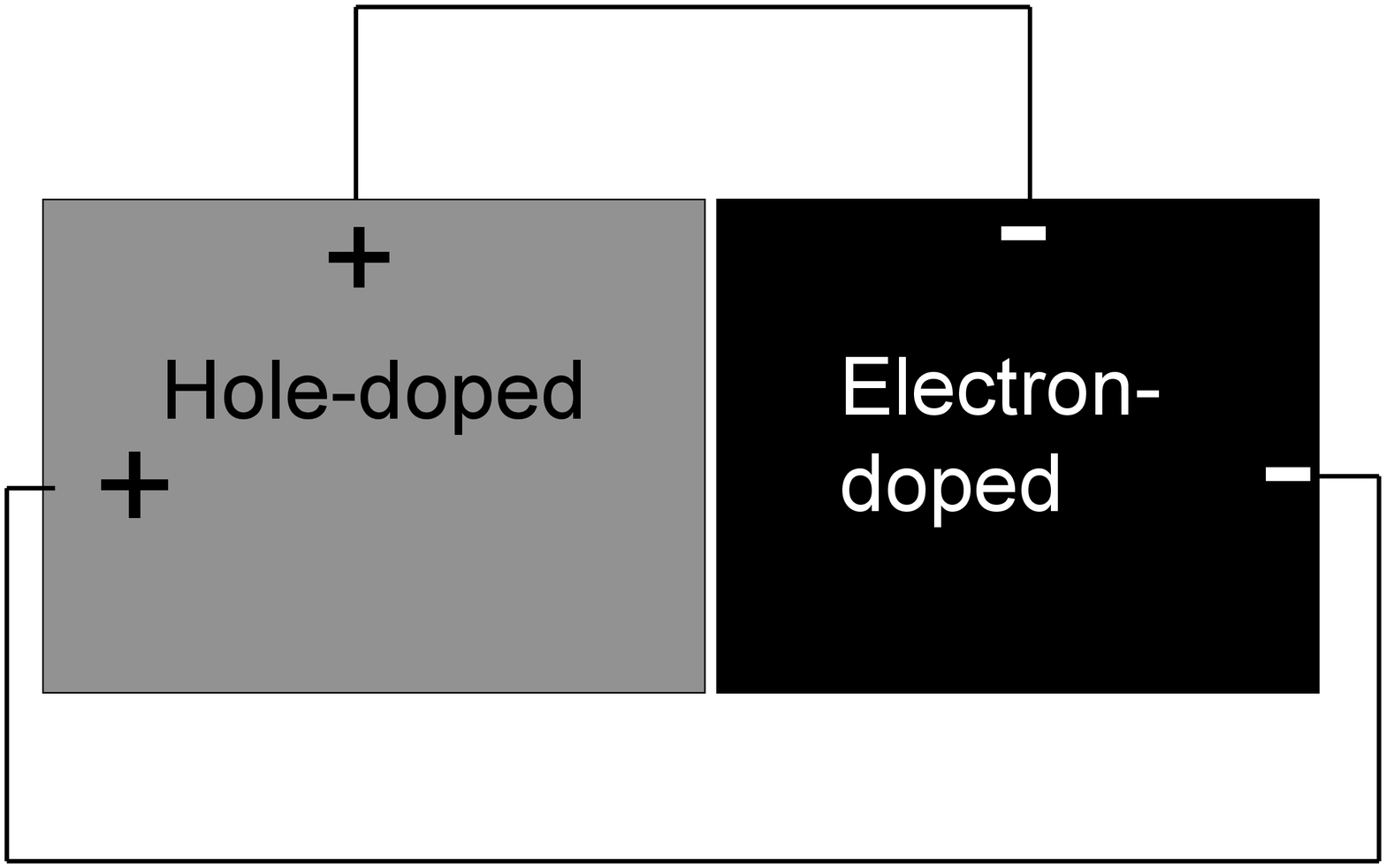} }
\caption{A schematic view of the tunneling geometry for the proposed
bicrystal experiments. Top: a c-axis orientation; bottom, an ab-plane
orientation with two possible lead orientations.}
\end{figure}
\newline
\noindent

A possible disadvantage of the proposed experiment is that it requires a
rather fine control over the interface properties. However, there is yet
another possibility of designing a two-junction experiment with a $\pi $
shift. This requires, however, a \textit{bi}crystal as shown in Fig. 3. We
propose to grow epitaxially a bicrystal of a hole-doped (Ba$_{1-x} $K$_{x}$Fe%
$_{2}$As$_{2})$ and an electron-doped (BaFe$_{2(1-x)}$Co$_{2x}$As$_{2})$
materials. As discussed above, the doping enhances the size and Fermi
velocity of the respective Fermi surfaces, and the conductance is dominated
by the hole or electron Fermi surface, correspondingly. The only remaining
problem is to ensure the proper phase coherence, that is, that the holes in
both crystals have the same phase, and the electrons the same, but opposite
to that of the holes.

In case of an epitaxial (coherent) interface the parallel wave vector, $%
k_{\parallel },$ is conserved through the interface, and the way to ensure
that the h-h and e-e currents are much larger than the e-h and h-e current
is to ensure that the overlap of the FS projections onto the interface plane
is maximal for the e-e and h-h overlaps, as opposed to the e-h overlap.
Obviously, this condition is satisfied in a bicrystal with a (100) interface
-- there is no e-h overlap at all, and the e-e and h-h overlaps are nearly
maximal possible. Unfortunately, growing an epitaxial (100) interface may be
very difficult.

On the other hand, growing a (001), or \textquotedblleft
c-axis\textquotedblright\ interface is much more natural. Let us consider
the FS overlaps in this case. Figure 4 plots the projections of the
calculated\cite{calc} Fermi surfaces of BaFe$_{1.6}$Co$_{0.4}$As$_{2}$ (dark
red) and Ba$_{0.6}$K$_{0.4}$Fe$_{2}$As$_{2}$ (light green). In this figure
the three dimensional Fermi surfaces have been telescoped onto the basal
plane, so that what one sees is the extent of the Fermi surface in the
planar direction across all wavevectors. The doping levels of $\pm 20\%$
were chosen because this is the \textquotedblleft
critical\textquotedblright\ spread at which the direct overlap of the e-FSs
nearly disappears. At any smaller spread there is either direct e-e overlap
or both e-e and h-h overlaps. Obviously, there is no e-h overlap and e-h
transport requires substantial nonconservation of the parallel momentum. 
\begin{figure}[h]
{\normalsize \includegraphics[width=7cm,angle=0]{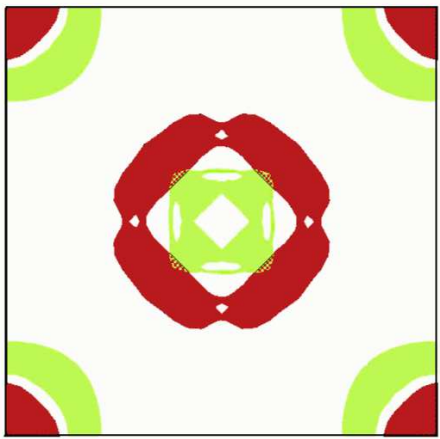} } \vspace{-0.7cm}
\caption{(Color online) A first-principles calculation of the ab-plane
projected three dimensional Fermi surfaces of Ba$_{0.6}$K$_{0.4}$Fe$_{2}$As$%
_{2}$ (green/gray) and BaFe$_{1.6}$Co$_{0.4}$As$_{2}$ (red/black).}
\end{figure}

In conclusion, we have proposed several phase-sensitive Josephson tests of
the ostensible $s_{\pm }$ order parameter symmetry in the superconducting
pnictides. The first design involves ab-plane corner junctions with angles
smaller than 90$^{0},$ the second either ab or ac 90$^{0}$ junctions,
prepared in such a way that one junction barrier is thin (specular) and the
other thick, and the third, probably the most promising one, uses
epitaxially grown hole- and electron-doped bicrystals in a \textquotedblleft
sandwich\textquotedblright\ orientation. We await the results of such
Josephson tunneling experiments with great interest.

It must be pointed out that there are several unknowns complicating
observation of the interferometric effect proposed. As opposed to d- or
p-wave pairing, the $\pi $ shift here is not a qualitative, symmetry
determined effect, but a quantitative one, based upon favorable relations
for tunneling probabilities for different bands. While we have taken into
account some major factors, accurate calculations of the said probabilities
are not possible. Interface properies may greatly affect them.

For these reasons the arguments given above should be considered in the
following light: if a $\pi $ phase shift between the electron and hole Fermi
surfaces is observed in any of the prposed geometries, this would be
extremely strong evidence for an $s_{\pm }$ state; unfortunately, the lack
of observation of such a shift in any given experiment cannot be taken as
similarly strong evidence against such a state.

We would like to acknowledge valuable discussions with R. Greene, I.
Takeuchi, and X. Zhang, particularly for suggesting the \textquotedblleft
sandwich\textquotedblright\ geometry.


\begin{thebibliography}{99}
\bibitem{harlingen} D.J. van Harlingen, Rev. Mod. Phys.  \textbf{67}, 515
(1995).

\bibitem{shen} A. Damascelli, Z. Hussain, and Z.X. Shen, Rev. Mod. Phys. 
\textbf{75}, 473 (2003).

\bibitem{wollman} D.A. Wollman, D.J. van Harlingen, W.C. Lee, D.M. Ginsberg,
and A.J. Leggett, Phys. Rev. Lett. \textbf{71}, 2134 (1994).

\bibitem{tsuei} C.C. Tsuei et al, Phys. Rev. Lett. \textbf{73}, 593 (1994).

\bibitem{geshkenbein} V.B. Geshkenbein, A.I. Larkin and A. Barone, Phys.
Rev. B \textbf{36}, 235 (1987).

\bibitem{rice} M. Sigrist and T.M. Rice, J. Phys. Soc. Jpn. \textbf{61},
4283 (1992).

\bibitem{nelson} K.D. Nelson et al., Science \textbf{306}, 1142 (2004).

\bibitem{kamihara} Y. Kamihara et al, J. Am. Chem. Soc. \textbf{130}, 3296
(2008).

\bibitem{mazin} I.I. Mazin, D.J. Singh, M.D. Johannes and M.H. Du, Phys.
Rev. Lett. \textbf{101}, 057003 (2008).

\bibitem{christianson} A.D. Christianson et al, arXiv:0807.3932.

\bibitem{demler_rmp} E. Demler, W. Hanke and S.C. Zhang, Rev. Mod. Phys. 
\textbf{76}, 909 (2004).

\bibitem{eremin_epl} M. M. Korshunov and I. Eremin, Phys. Rev. B \textbf{78}%
, 140509(R), (2008).

\bibitem{scalapino} T.A. Maier and D.J. Scalapino, Phys. Rev. B \textbf{78},
020514 (2008).

\bibitem{zhou} Y.-R. Zhou et al, arXiv:0812.3295.

\bibitem{zhang} X. Zhang et al, arXiv:0812.3605.










\bibitem{mazin_epl} I.I. Mazin, Europhys. Lett. \textbf{55}, 404 (2001).

\bibitem{MazinPRL} I.I. Mazin, Phys. Rev. Lett. \textbf{83}, 1427 (1999).

\bibitem{landau} L.D. Landau and E.M. Lifshits, \textit{Quantum Mechanics},
(Pergamon Press, Oxford, New York), 1977, section 2.3, Eq. 25.

\bibitem{calc} We used the linear augmented plane wave method (LAPW) in the
virtual crystal approximation, as discussed in Ref. \cite{mazin}.
\end{thebibliography}
\end{document}